\documentclass[11pt,a4paper]{article} 
\usepackage{jcappub} 

\usepackage{amsfonts}
\usepackage{varioref}
\usepackage{comment}

\usepackage{amssymb}
\usepackage{latexsym}
\usepackage{graphicx,subfigure,graphics,epsfig,amsmath,amssymb}
\usepackage{hyperref}
\usepackage{bm}
\usepackage{ctable}

\newcommand{\be}{\begin{equation}}
\newcommand{\ee}{\end{equation}}
\newcommand{\beqa}{\begin{eqnarray}}
\newcommand{\eeqa}{\end{eqnarray}}

\newcommand{\AS}{A_{\rm S}^2}
\newcommand{\AT}{A_{\rm T}^2}
\newcommand{\nS}{n_{\rm S}}
\newcommand{\nT}{n_{\rm T}}
\newcommand{\run}{d\nS/ d \ln k}

\newcommand{\impc}{\ensuremath{{\rm\,Mpc}^{-1}}}

\newcommand{\simlt}{\lower.5ex\hbox{$\; \buildrel < \over \sim \;$}}
\newcommand{\simgt}{\lower.5ex\hbox{$\; \buildrel > \over \sim \;$}}

\newcommand{\iMpc}{\impc}

\newcommand{\apjs}{{Astrophys.~J.~Supp. }}

\newcommand{\mnras}{{Mon.~Not.~R.~Astron.~Soc.}}

\usepackage{natbib}

\begin{document}

\title{On the prior dependence of constraints on
  the tensor-to-scalar ratio}
\author[a,b]{Marina Cort\^{e}s,} 
\author[c]{Andrew R. Liddle,}
\author[d]{and David Parkinson}
\affiliation[a]{Berkeley Lab, Berkeley, CA 94720, USA}
\affiliation[b]{African Institute for Mathematical Sciences,
  Muizenberg 7945, Cape Town, South Africa} 
\affiliation[c]{Astronomy Centre, University of Sussex, Falmer,  
Brighton BN1 9QH, UK}
\affiliation[d]{School of Mathematics and Physics, University of Queensland,
  Brisbane, QLD 4072, Australia} 
  
 \emailAdd{mcortes@lbl.gov}
 \emailAdd{a.liddle@sussex.ac.uk}
 \emailAdd{d.parkinson@uq.edu.au}

\date{\today}

\abstract{
We investigate the prior dependence of constraints on cosmic tensor
perturbations. Commonly imposed is the strong prior of the
single-field inflationary consistency equation, relating the tensor
spectral index $\nT$ to the tensor-to-scalar ratio $r$. Dropping it
leads to significantly different constraints on $\nT$, with both
positive and negative values allowed with comparable likelihood, and
substantially increases the upper limit on $r$ on scales $k=0.01
\iMpc$ to $0.05 \impc$, by a factor of ten or more. Even if the
consistency equation is adopted, a uniform prior on $r$ on one scale
does not correspond to a uniform one on another; constraints therefore
depend on the pivot scale chosen. We assess the size of this effect
and determine the optimal scale for constraining the tensor amplitude,
both with and without the consistency relation.}


\maketitle

\section{Introduction} 

The inflationary proposal \cite{inf} is a compelling scenario for the
origin of primordial density fluctuations. Its prescription for the
spectrum of density perturbations has passed the scrutiny of high
precision measurements of the Cosmic Microwave Background (CMB), 
being
in agreement with small deviations from a purely scale-invariant
spectrum \cite{wmap7}.  In all models of inflation, a small but
potentially non-negligible stochastic background of primordial
gravitational waves is also generated alongside the density
fluctuations. This prediction remains to be confirmed, with only an
upper bound on the density of gravitational waves so far.

Although there are currently no useful theoretical lower bounds on the
tensor amplitude from inflation models (since we have classes of
models that predict undetectable levels of tensors \cite{string}),
determining the amplitude of gravitational waves or, failing that,
imposing a stringent upper bound is still formally a requirement for
completion of the underlying physical picture behind the inflation
mechanism. The amplitude of gravitational waves determines the energy
scale at which inflation took place, shedding light on the particle
physics sector in those regimes, and reveals the total excursion of
the scalar field during inflation. This provides information on
symmetries that may have been broken at the time, thereby probing
energy scales many orders of magnitude larger than those at the Large
Hadron Collider.

One of the predictions of the simplest models of inflation is that the
spectrum of tensor perturbations depends on that of scalars. The
spectra have a common origin in the same function, the potential of
the single scalar field, and are therefore related.  This prediction
characterizes the simplest models of inflation (single field and
slowly rolling), and yields a set of consistency relations all valid to
a given order in the slow-roll regime \cite{cortes_liddle}. These
models give a good description of current data \cite{wmap7}.

Nevertheless, other assumptions on the nature of tensor perturbations
are possible.  More general classes of inflation models include those
with multiple scalar fields \cite{two-field}, where the consistency
equation becomes an inequality \cite{sasaki_stewart}, and those with
deviations from the slow-roll mechanism.  In addition there are models
in which density fluctuations are seeded through a physical mechanism
making use of the known duality between expanding and contracting
cosmologies.  In collapse-type models the rapidly expanding horizon in
an inflationary era is associated to contraction in a matter-dominated
era \cite{prebigbang}. In ekpyrotic models the slow contraction takes
place during domination by a stiff fluid with equation of state $w>1$
\cite{ekp}. Alternatively, tensor perturbations might be seeded by
cosmic defects \cite{defects}, or one might simply ask what can be
learnt from observations without imposing specific theoretical
preconceptions.  There are therefore a variety of possible assumptions
as to how the tensor perturbations might behave, and some examples have 
been investigated in the literature.

Finelli, Rianna, and Mandolesi \cite{finelli} tested the validity of the 
consistency 
equation by sampling freely from the plane $16\,\epsilon-r$. They probed 
tensors at a scale close to the scalar pivot, $k=0.01\impc$, and not at 
$k=0.002\impc$. 
Camerini et al \cite{camerini} dropped the assumption of single field inflation 
and searched at $k=0.002\impc$ for compatibility with blue tilted tensors in 
current data. The applied prior on $\nT$ was $ -1 < \nT < 20$.
This apparently strongly conservative prior still misses a large fraction of 
allowed red tilted models at $\nT<-1$. Also they probed solely at 
$k=0.002\impc$ and thus didn't detect the weakening of constraints as we 
move towards smaller scales.
Valkenburg, Krauss, and Hamann \cite{valkenburg} considered a prior 
density 
on $r$ which differs from the common choice of a flat prior. They select for 
a 
theoretically motivated uniform prior on the energy scale of inflation. They 
also 
probe at a single scale, $k=0.002\impc$. Even assuming the 
consistency equation they detect that constraints on $r$ reflect this change 
on 
the tensor prior.
Zhao, Baskaran and Zang, \cite{zhao} searched for the best multipole to 
probe 
tensors at. Here they dropped the consistency equation assumption and 
probed tensors at more than one scale. They then draw forecasts under 
this 
setup, for a variety of ground- and space-based upcoming B-mode 
experiments. 
The chosen fiducial model had small tensor amplitude and as result they 
don't 
detect the large variation allowed at the short wavelengths.
In Ref.~\cite{gjerlow_elgaroy} Gjerl\o w and Elgar\o y relax the consistency 
equation relation but retain the assumption of a nearly scale invariant 
tensor 
spectrum $\nT\sim\nS-1$.
Finally Powell \cite{powell} investigates the sensitivity of upcoming B-mode 
experiments for detecting the tensor tilt, both for blue and red $\nT$.

In our work we allow for full variability in these sets of assumptions, and 
make a 
systematic exploration of how constraints on tensor fluctuations respond to 
changes of prior.
We relax the imposition of the simplest inflation models, vary the 
cosmological 
scale probed, and study different prior densities on both $r$ and $\nT$. 
This 
permits us to unveil the underlying variation of constraints on tensor 
parameters, that result from shifts in one's set of priors, and we observe a 
larger range of behaviour of the tensor spectra than any of the studies in 
the literature so far.


\section{Methodology}

We assume throughout a flat $\Lambda$CDM model and 
parameterize our set of primordial spectra as
\beqa
\AS(k)&=&\AS(k_0) k^{\nS(k_0)-1+\alpha(k_0) \ln k/k_0}\,,\\
\AT(k)&=&\AT(k_0) k^{\nT}\,,
\eeqa
where $\alpha\equiv \run$, and $k_0$ is the pivot scale where
observables are specified. We  define the ratio of tensor to 
scalar amplitude of perturbations as 
\be
r(k) \equiv 16 \frac{\AT(k)}{\AS(k)} \,.\label{r}
\ee

The usual assumption when fitting primordial fluctuations for
parameter estimation is the imposition of the first consistency
relation between the scalar and tensor power spectra,
\begin{equation}\label{ce}
r = - 8 \nT \,.
\end{equation}
For inflationary models the first and second derivatives of the scalar
field potential are usually expressed in terms of the slow-roll
parameters \cite{LL92},
\begin{equation}\label{epseta}
\epsilon =\frac{m_{{\rm
      Pl}}^2}{16\pi}\left(\frac{V'}{V}\right)^2 \quad ; \quad
\eta = \frac{m_{{\rm Pl}}^2}{8\pi}\, \frac{V''}{V}\,,
\end{equation}
where prime indicates $d/d\phi$.
In the context of the slow-roll expansion and a single field, we can
write the observables in terms of the derivatives of the potential
\cite{LL92} 
\begin{eqnarray}\label{nSnT}
\nS-1&=&-6 \epsilon + 2 \eta \,; \label{nS} \\
\nT&=& -2 \epsilon \,; \label{nT}\\
r & = & 16\epsilon \,,
\end{eqnarray}
from which the first consistency relation immediately follows.  A
consequence of the consistency equation Eq.~(\ref{ce}) is to enforce
the amplitude of tensors to be a decreasing function of wavenumber, since
$r$ is by construction always positive.

In multi-field models, tensor perturbations are still given by the
usual formula but extra scalar perturbations can be generated by
conversion of isocurvature perturbations in the additional degrees of
freedom. The consistency equation then weakens to an inequality, $r <
-8\nT$ with $\nT$ still constrained to be negative
\cite{sasaki_stewart}.  Other models may give yet further variation,
e.g.\ collapsing Universe inflation models such as the pre-big-bang
models \cite{prebigbang} give a positive tensor spectral index.

In this paper we carry out an extensive exploration of the prior
dependence of constraints on the tensor-to-scalar ratio. The prior
dependence takes two forms: (a) restrictions on the parts of the
$r$--$\nT$ plane that are considered, e.g.\ by enforcing the above
consistency equation or inequality, and (b) the prior probability
distribution adopted on the permitted parameter region. For the
latter, commonly a uniform prior is chosen at whichever `pivot' scale
has been selected to specify the parameters, but this picks out a
special scale for which this is being assumed true. A related question
is to ask on what scale the tensor-to-scalar ratio is actually best
constrained by a given dataset, given prior assumptions.

In order to explore these dependencies, we will consider three
different cases:
\begin{enumerate}
\item No consistency relation, i.e.\ $r$ and $\nT$ able to vary
  independently including positive $\nT$.
\item Single-field consistency equation imposed, i.e\ $r = -8\nT$.
\item Multi-field consistency inequality imposed, i.e.\ $r \leq -8\nT$.
\end{enumerate}
In each case we need to consider the effect of imposing priors at
different scales; we choose $0.002\iMpc$ as used in WMAP papers
\cite{wmap5,wmap7} and $k=0.02 \iMpc$ which is about the scale at
which scalar perturbations are optimally constrained
\cite{cortes_liddle_mukherjee}.

\section{Constraints on the tensor amplitude}

\ctable[ 
caption = Uniform priors assumed in this article., 
label	= prior,
]{llcc}
{  
\tnote[a]{at scale $k=0.002 \impc$} 
\tnote[b]{at scale $k=0.02 \impc$}
}
{
\ML
& &lower &upper \\	
\FL
Physical baryon density 		& $\Omega_{\rm b}h^2$ 		&  0.005	
&0.1\\
Physical dark matter density 	& $\Omega_{\rm cdm}h^2$ 	&  0.01	
&0.99\\
Sound horizon 				& $\theta$					&  0.5	
&10\\
Optical depth 				& $\tau$ 					&  0.01	
&0.8\\
Scalar amplitude			& $\log (10^{10} \AS)$ 		&  2.7	&4\\
Scalar index 				& $\nS$ 					&  0.5	
&1.5\\
Scalar running 				& $\run$ 					&$-0.2$	
&0.2\\
SZ-amplitude				& $A_{\rm SZ}$ 			&  0		&2\\
Tensor-to-scalar ratio 		& $r$					&0		&
$1\tmark[a]$\\
						&						&0		&
$16\tmark[b]$\\
Tensor spectral index 		& $\nT$					&$-4$	&
$4\tmark[a]$\\
						&						&$-4$	&
$8\tmark[b]$\\
\hline
}

To probe parameter space we use the Markov Chain Monte Carlo method 
as
implemented in the CosmoMC package \cite{cosmomc}, and assume our
cosmology can be described by the set of parameters in
Table~\ref{prior} where we specify the uniform priors imposed.  All
our MCMC runs use the combination of datasets from WMAP-7 year
\cite{wmap7}, matter power spectrum measurements from SDSS-DR7
\cite{dr7}, and $H_0$ from the Hubble Science Telescope (HST)
\cite{hst}. Though the non-CMB experiments (SDSS and HST) cannot
detect the tensor perturbations directly, these extra data are
necessary to break some of the parameter degeneracies between $r$ and
other cosmological parameters.

\begin{figure} [t]
\centering
$\begin{array}{cc}
\includegraphics[width= 0.5\textwidth]
{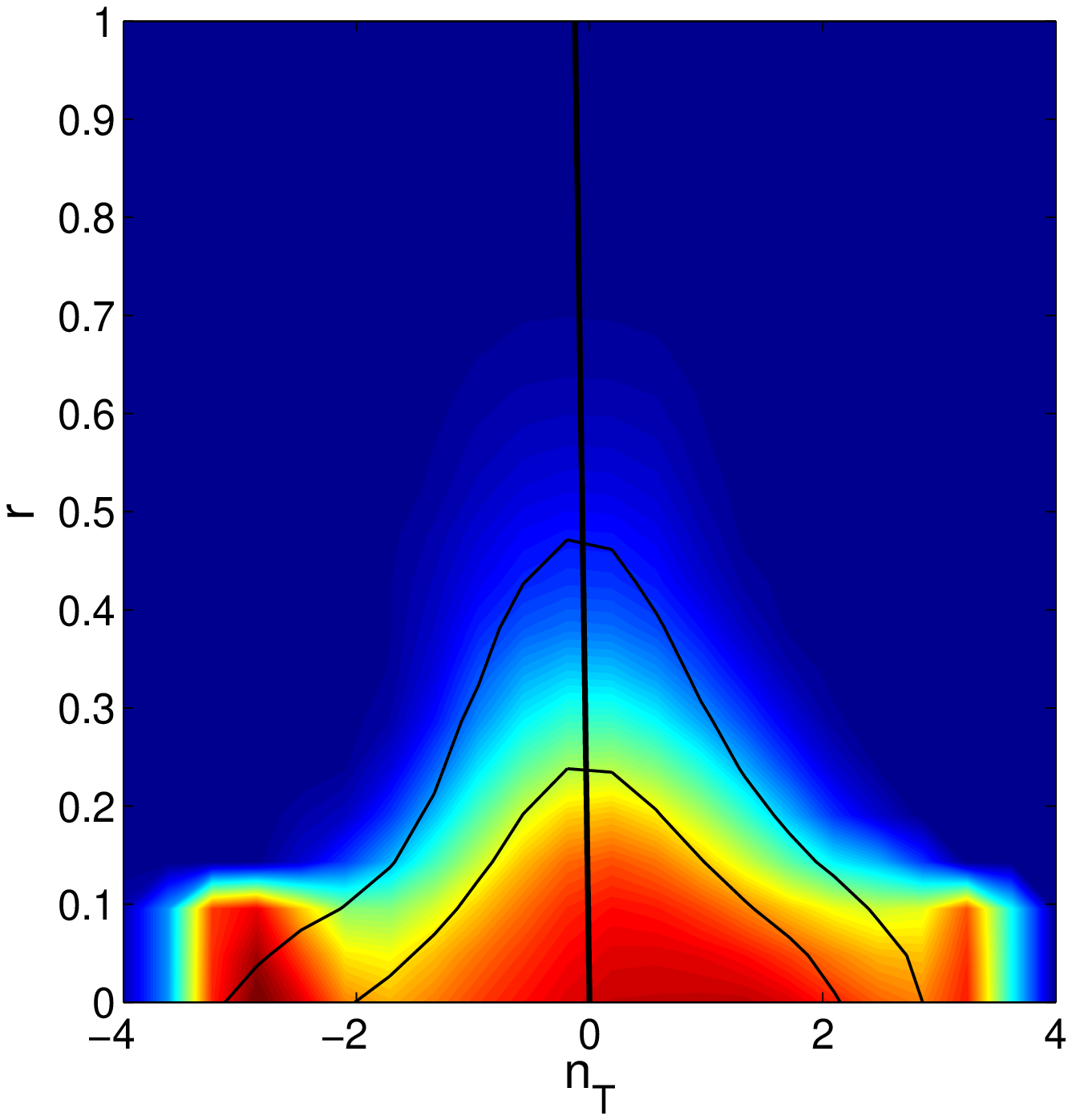}&
\includegraphics[width= 0.5\textwidth]
{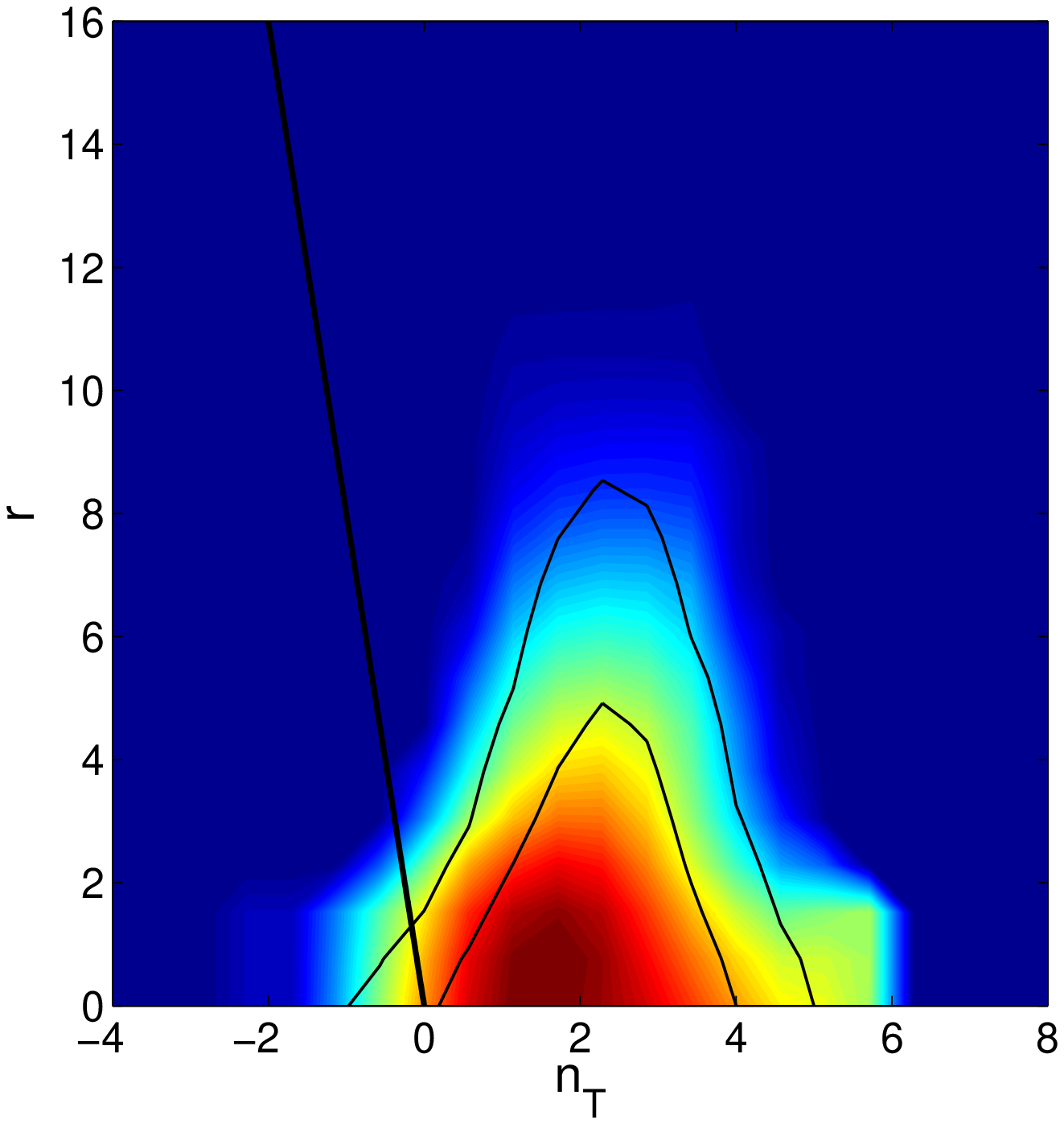}\\
\end{array}$
\caption{The $\nT$--$r$ two-dimensional distributions obtained at 
different scales, without imposition of the consistency equation, 
and uniform priors over the regions plotted. \textbf{Left panel}: 
constraints obtained at $k=0.002\impc$; \textbf{right panel}: 
constraints at scale $k=0.02\impc$. The colour scale represents 
the mean  likelihood of samples, going to higher likelihood 
towards redder regions, and the contours denote the marginalized 
distributions at  68\% and 95\% confidence limits. The straight line shows 
the consistency equation.}
\label{ntr0.002_ntr0.02} 
\end{figure}

\subsection{No consistency relation}

We begin by discussing the most general case where $r$ and $\nT$ vary
freely, limited only by priors that we have chosen to be so wide that
the posterior distributions are limited by the data, weak though they
are in constraining the tensors, particularly their scale
dependence. As well as displaying these `model independent'
constraints, we use this case to explore the effect of pivot scale
transformations.

Figure~\ref{ntr0.002_ntr0.02} (left panel) shows combined constraints on $
\nT
$ and $r$,
with uniform priors imposed at $k=0.002 \iMpc$; similar results were 
obtained by Camerini \textit{et al} \cite{camerini}. The consistency
equation would have restricted models to the line, and obviously
dropping this assumption greatly increases the parameter space that is
available. As a byproduct, this figure shows how far we are from a
meaningful test of the consistency relation, which would require
observational constraints tightly associated to a point on the
consistency equation line or excluding it entirely.

We find that the data do constrain $\nT$, albeit very weakly, both
from above and from below. The constraints weaken as $r$ becomes
small, because $\AT$ is very small there and $\nT$ is irrelevant if
the amplitude is too small to detect on all scales probed.  Large very
positive $\nT$ would eventually contribute to the temperature modes at
large $\ell$ and conversely large negative $\nT$ would disrupt the
Sachs--Wolfe effect beyond cosmic variance.

Note that the data allow the spectral tilt of tensor perturbations to
be positive, contrary to standard inflationary predictions.  Indeed 
the one-dimensional marginalized confidence limits on
$\nT$ are approximately symmetric about zero: 
\be 
n_{\rm T} (k=0.002\impc)=0.0^{+1.1\, +2.0}_{-0.9\,-2.0} \,,
\ee
where uncertainties quoted are 68\% and 95\%.

We now wish to investigate the constraints if the tensors are
specified at a shorter scale, $k=0.02 \iMpc$. The right panel of Figure~\ref
{ntr0.002_ntr0.02} 
shows the $\nT$--$r$ plane when we change the pivot scale to $k=0.02
\impc$ and otherwise maintain the same setup, i.e.\ the same datasets
and priors on remaining cosmology. The principal difference is that
the priors on parameters are now assumed uniform at this scale; we
also widened the ranges of $r$ and $\nT$ in order to be able to reach
regions constrained by the data. Motivated by the fact that $\epsilon
=1$ signals the end of slow-roll inflation, and that $r=16 \epsilon$
from the consistency equation, we allow for $0<r<16$ and choose the
prior on $\nT$ such that 95\% limits are well within the prior range
$-4< \nT<8$.

\begin{figure} [t]
\centering
$\begin{array}{cc}
\includegraphics[width= 0.5\textwidth]{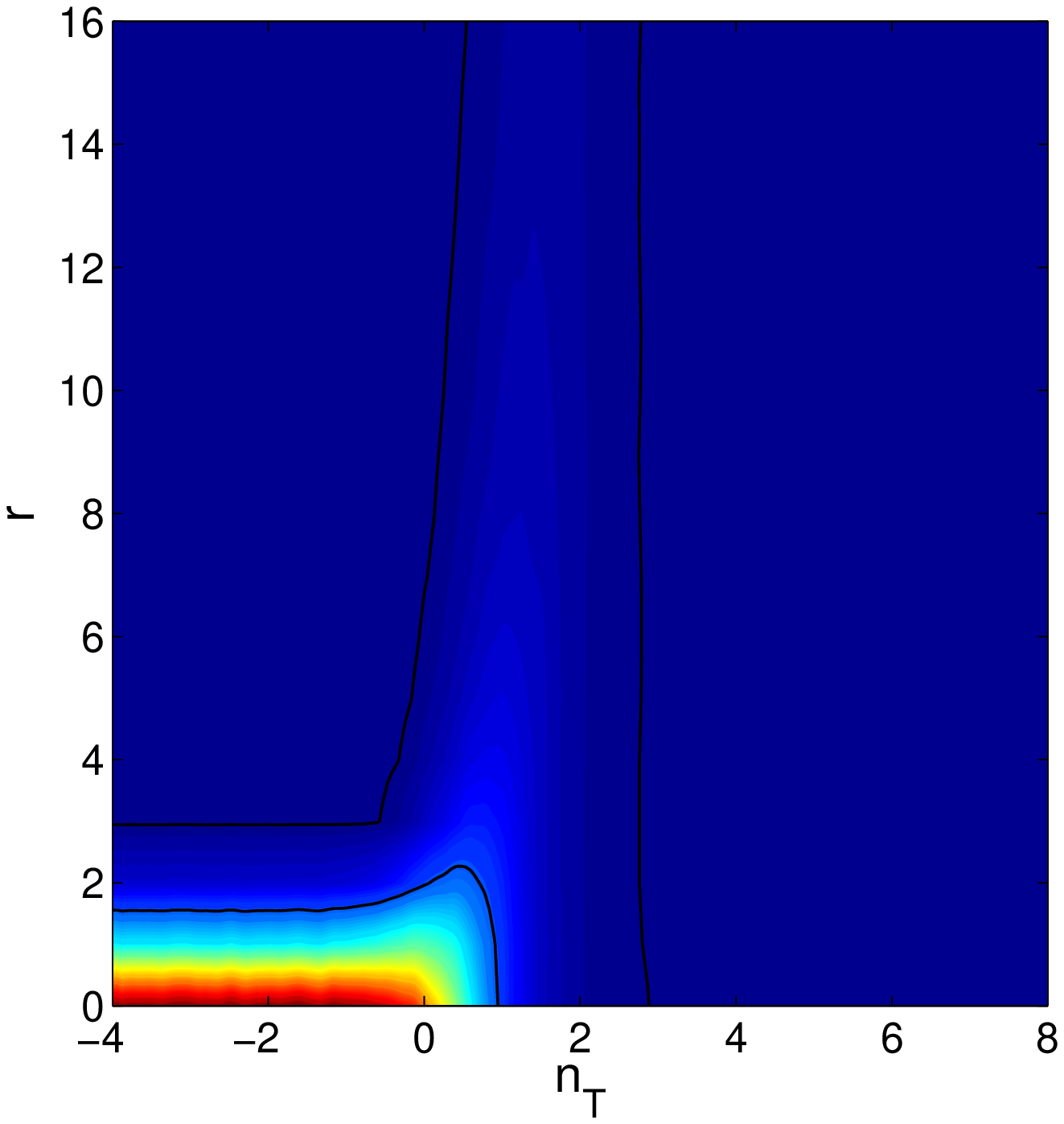}&
\includegraphics[width= 0.5\textwidth]{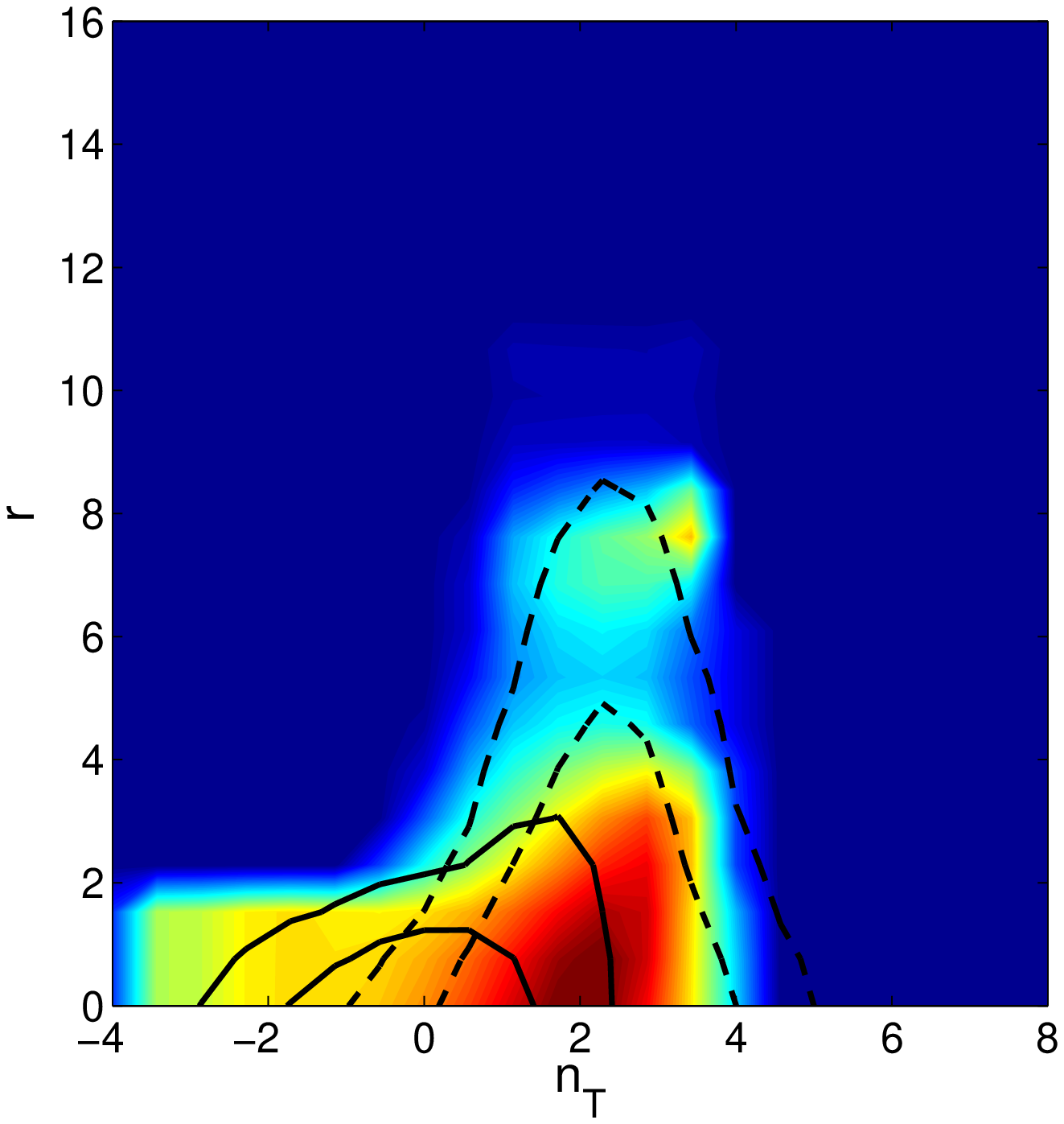}\\
\end{array}$
\caption{\textbf{Left panel:} The induced prior at $k=0.02\impc$ 
from a uniform prior at $k = 0.002\impc$. \textbf{Right Panel:} 
superposition of two-dimensional distributions of $\nT$--$r$ at  
$k=0.02\impc$. The dashed lines represent the marginalized
distributions of $\nT$--$r$ obtained by running a chain with uniform
prior $k=0.02\impc$. The solid lines and colour code represent the
distributions of $\nT$--$r$ obtained in a chain run with uniform
prior at $k=0.002\impc$ and shifted to $k=0.02\impc$.}
\label{ntr_prior_ovrl} 
\end{figure}

One might have expected that at least the constraints on $\nT$ would
have been unchanged, since it is unaffected by a transformation of
pivot scale. However we see that this is not the case; the
distribution is shifted significantly to the right. The reason is that
a uniform prior on one scale does not transform to a uniform prior on
another, and because the tensors are so weakly constrained (at least
in absence of consistency relations) the effect of this is quite
dramatic. Figure \ref{ntr_prior_ovrl} (left panel) shows the {\em prior} 
induced 
on
parameters at $0.02 \iMpc$ by a uniform prior at $k=0.002 \iMpc$. We
stress there is no data at all in this figure; we just draw uniform
points on one scale and analytically transform $r$ to the second scale
($\nT$ does not change). As we see, the resulting prior is extremely
non-uniform. Note that it is the region where the prior is compressed
that leads to a high prior (since the fraction of the prior in each
interval of $\nT$ is constant).

The consequence is shown in the right panel of Figure~\ref{ntr_prior_ovrl}, 
which overlays
the chain run at $k=0.02\impc$ with the chain run at the pivot scale
$k=0.002\impc$ and transposed to the smaller scale $k=0.02\impc$.  
The
dashed lines are the posterior assuming a uniform prior, matching
Fig.~\ref{ntr0.002_ntr0.02} (right panel), and hence also indicate the 
likelihood. The
posterior of the transformed chain is obtained by multiplying this
likelihood with the transformed prior of Fig.~\ref{ntr_prior_ovrl} (left panel), 
which
shifts the preferred region downwards and leftwards yielding the solid
lines. The choice of scale to impose the uniform prior clearly
substantially modifies the constraints on each parameter.

\begin{figure}[t]
\begin{center}
\includegraphics[width= 0.7\textwidth]{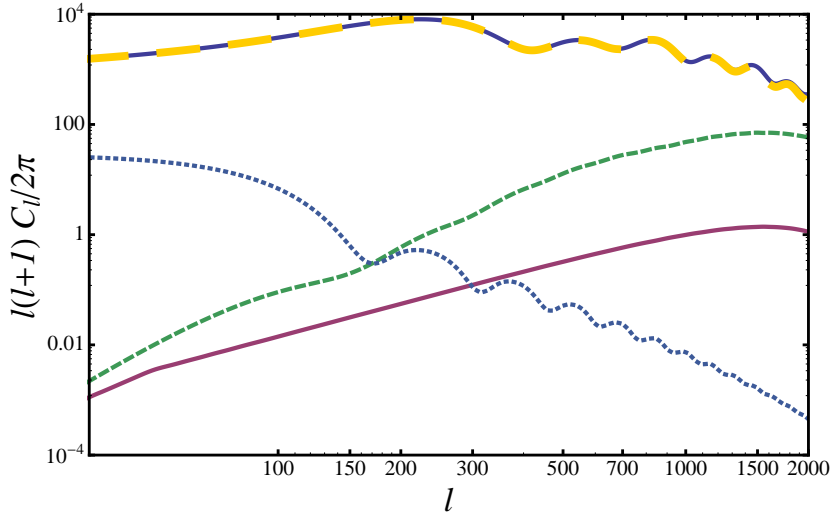}
\caption{Angular power spectra for a model selected from the
  2-dimensional $\nT-r$ distribution of Fig.~\ref{ntr0.002_ntr0.02} (right 
panel), 
allowed at
  95\% confidence level, and comparison with the single-field
  inflationary prediction. Parameter values are $\nT=5$, $\nS=0.956$,
  $r=0.05$ and $\run=-0.009$, all specified at scale
  $k=0.02\impc$. Solid blue and pink lines show TT and BB total power
  modes. Yellow long-dashed shows the scalar contribution to TT power,
  and green short-dashed the tensor contribution to TT. Blue dotted
  line shows the corresponding BB total power for the corresponding
  single-field model that would have been obtained by imposition of
  the consistency equation.}
\label{camb_large_nT} 
\end{center}
\end{figure}

Note that on this shorter scale the constraint on $r$ is considerably
weakened with the $2\sigma$ constraints increased by a factor of more
than ten (see below in Eq.~(\ref{nTrconstr0.02})). The primordial
amplitude of tensor fluctuations is allowed to be as large as six
times that of scalar fluctuations, and steeply-rising tensor spectra
of tilt $\nT=5$ are allowed at the $95\%$ level. We show the TT and BB
CMB spectra for such a model in Fig.~\ref{camb_large_nT}.
Even at $n_{{\rm T}}=0$, corresponding to scale-invariant tensors, there is 
a significant change in the upper limit on $r$ due to this change in prior, 
investigated in more detail later in this section.

The 1D (95\%) marginalized limits on $\nT$ and $r$ 
at the two scales are: 
\beqa\label{nTrconstr0.02}
-2.0 < &\nT& < 2.0\,, \quad r < 0.35 \quad {\rm at}\quad k=0.002\impc\,,
\nonumber\\
0.76 < &\nT& < 3.6\,,  \quad r < 6.6 \quad {\rm at}\quad k=0.02\impc
\eeqa
which reveals a significant shift towards more positive values in the
preferred $\nT$ at smaller scales, with the best fit changing by about
$1\sigma$. At this scale the interesting region for inflation,
$-1\lesssim \nT<0$, lies around the $1\sigma$ lower limit.  Naive
examination of these contours would lead to the conclusion that data
are favouring a larger value of $\nT$ at the smaller scale, and give
evidence of non-zero running of the tensor spectral index, given the
stark difference in confidence regions.  However, none of the models
fitting the data in the left panel of Fig.~\ref{ntr0.002_ntr0.02} fail the 
prior imposed in the right panel, i.e.\ no model with $r$, $\nT$ at 
$k=0.002\impc$ (left panel) is outside the confidence limits at 
$k=0.02\impc$ (right panel in the same figure). This means that the same
models are represented in the two distributions. It is rather the
effect of change in prior, i.e.\ the change in sampling at the new
scale, which is causing the change in the posterior.  This is true for
any parameter that transforms under cosmological scale, but is more
relevant if in addition the parameter is poorly constrained as in the
case of tensor spectra.

\subsection{With consistency relations}

\begin{figure*} [t]
\centering
$\begin{array}{ccc}
\includegraphics[width= 0.3\textwidth]{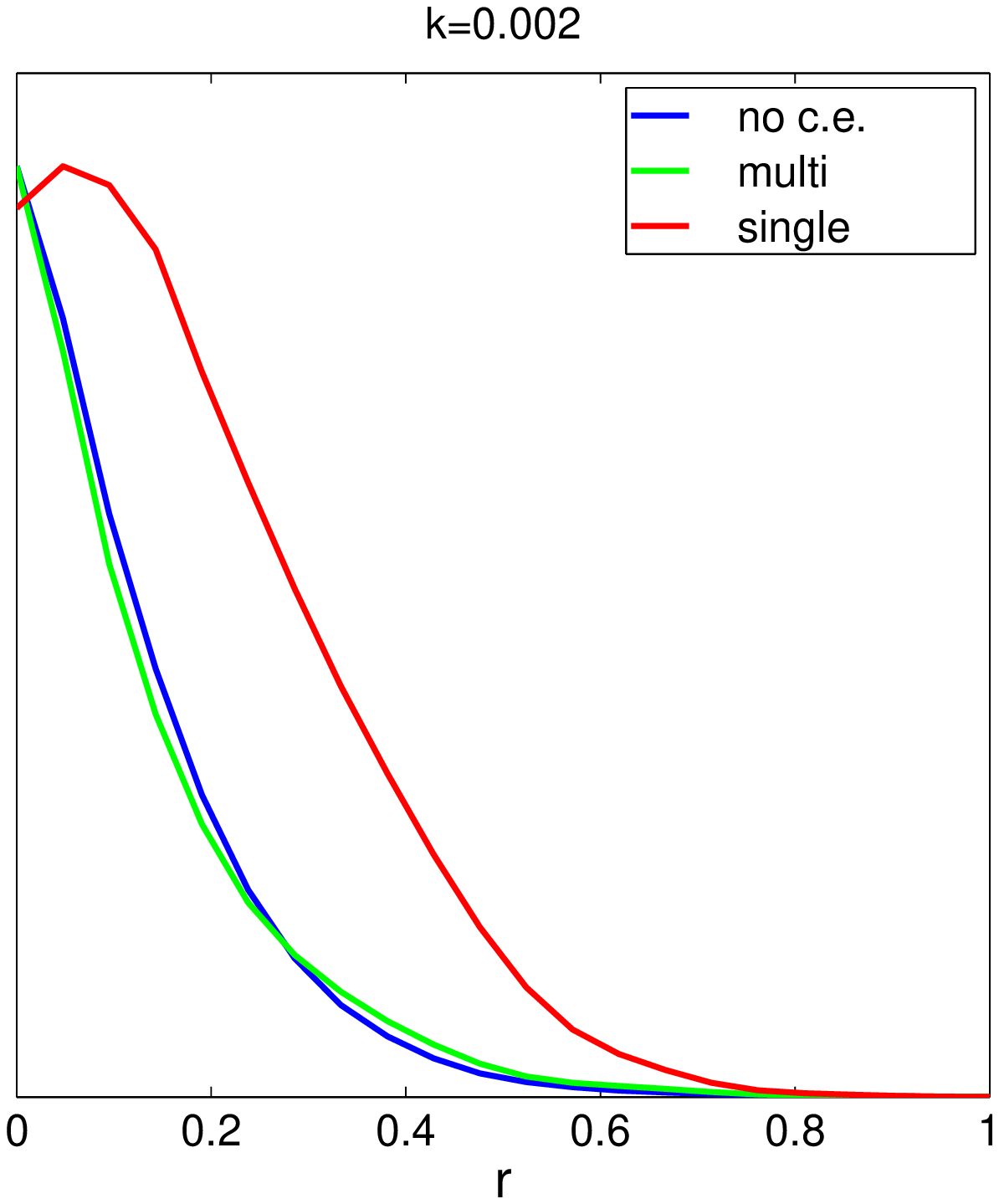}& 
\includegraphics[width= 0.3\textwidth]{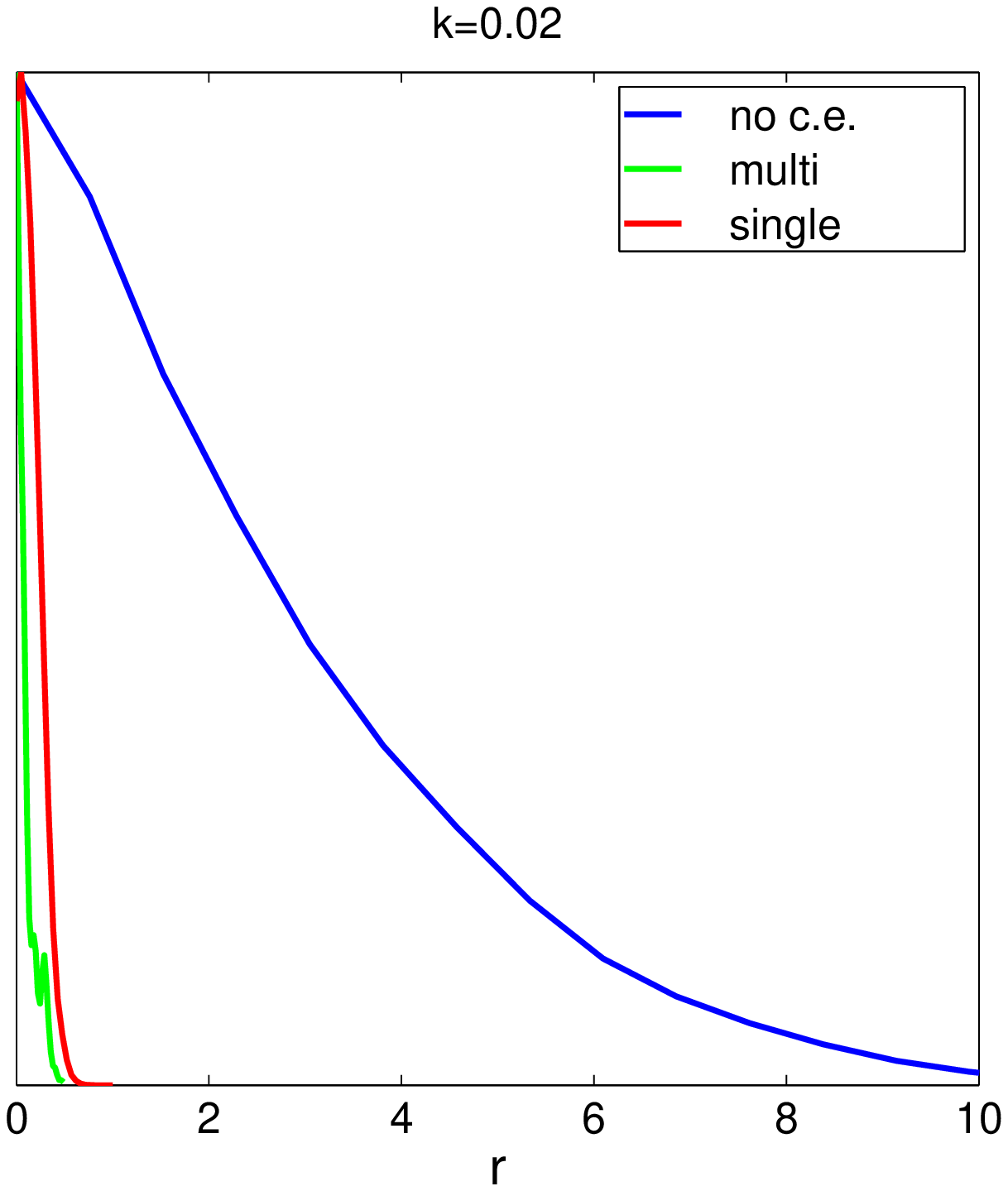}& 
\includegraphics[width= 0.3\textwidth]
{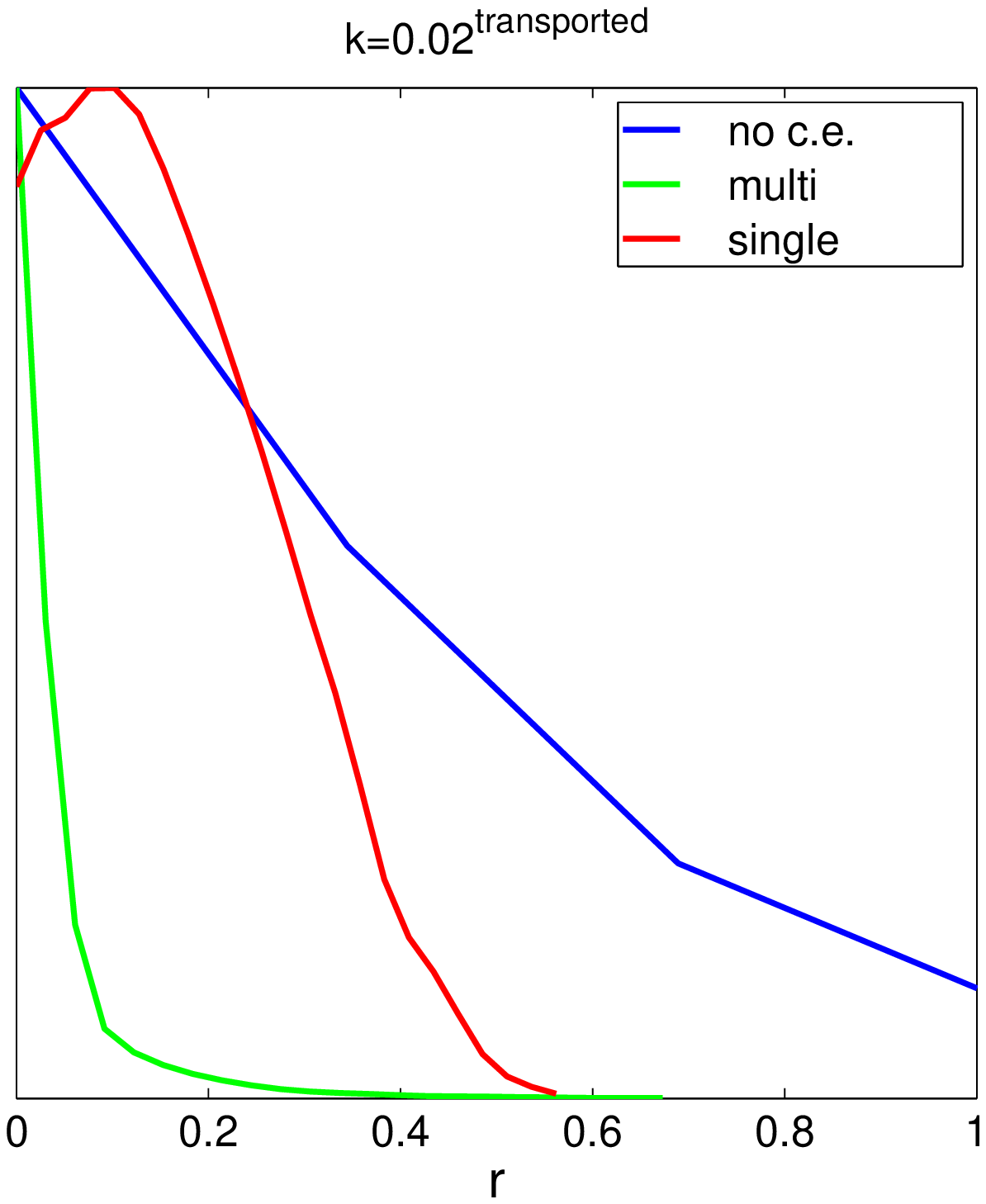}\\ 
\end{array}$
$\begin{array}{ccc}
\includegraphics[width= 0.3\textwidth]
{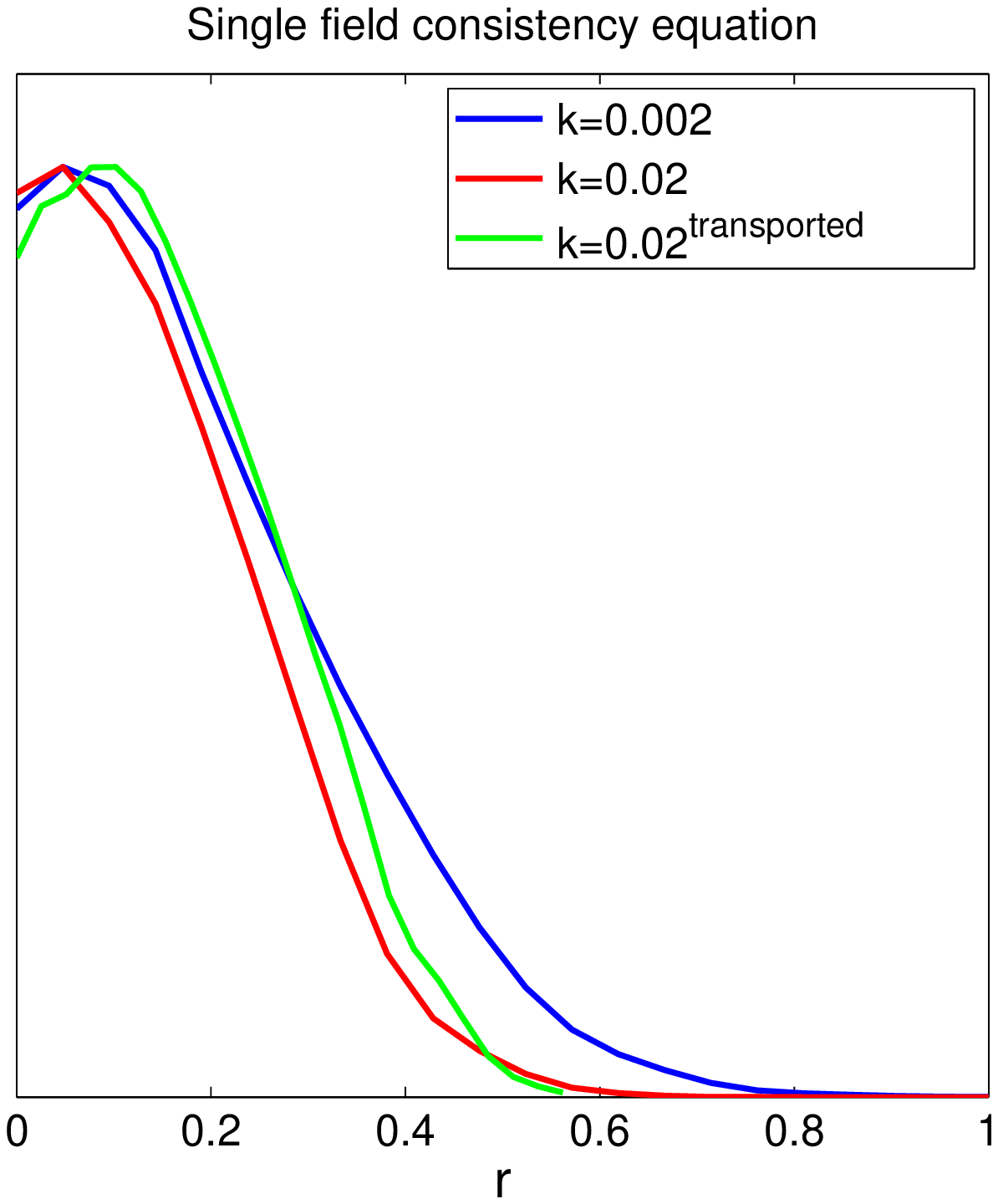}&
\includegraphics[width= 0.3\textwidth]
{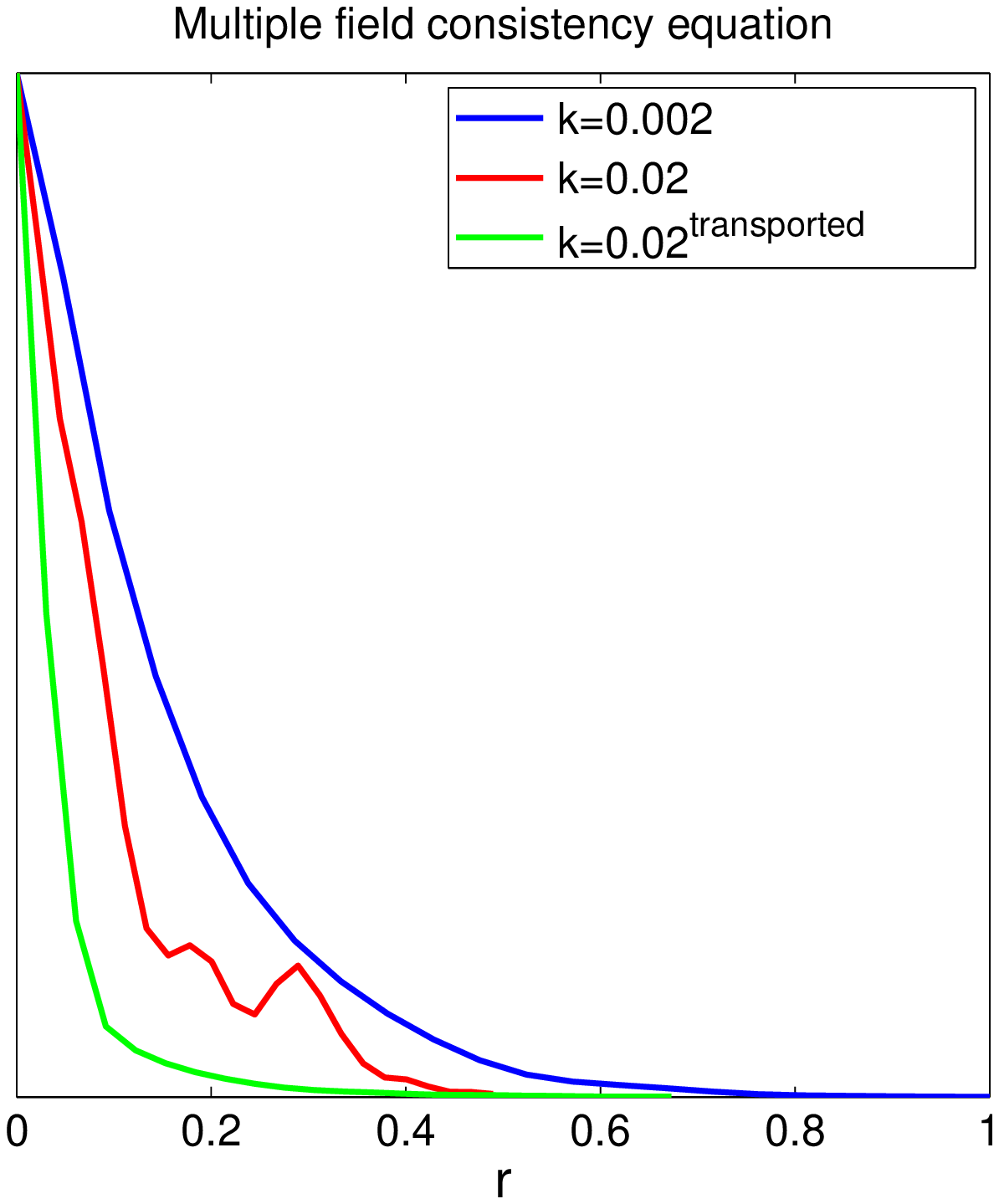}&
\includegraphics[width= 0.3\textwidth]
{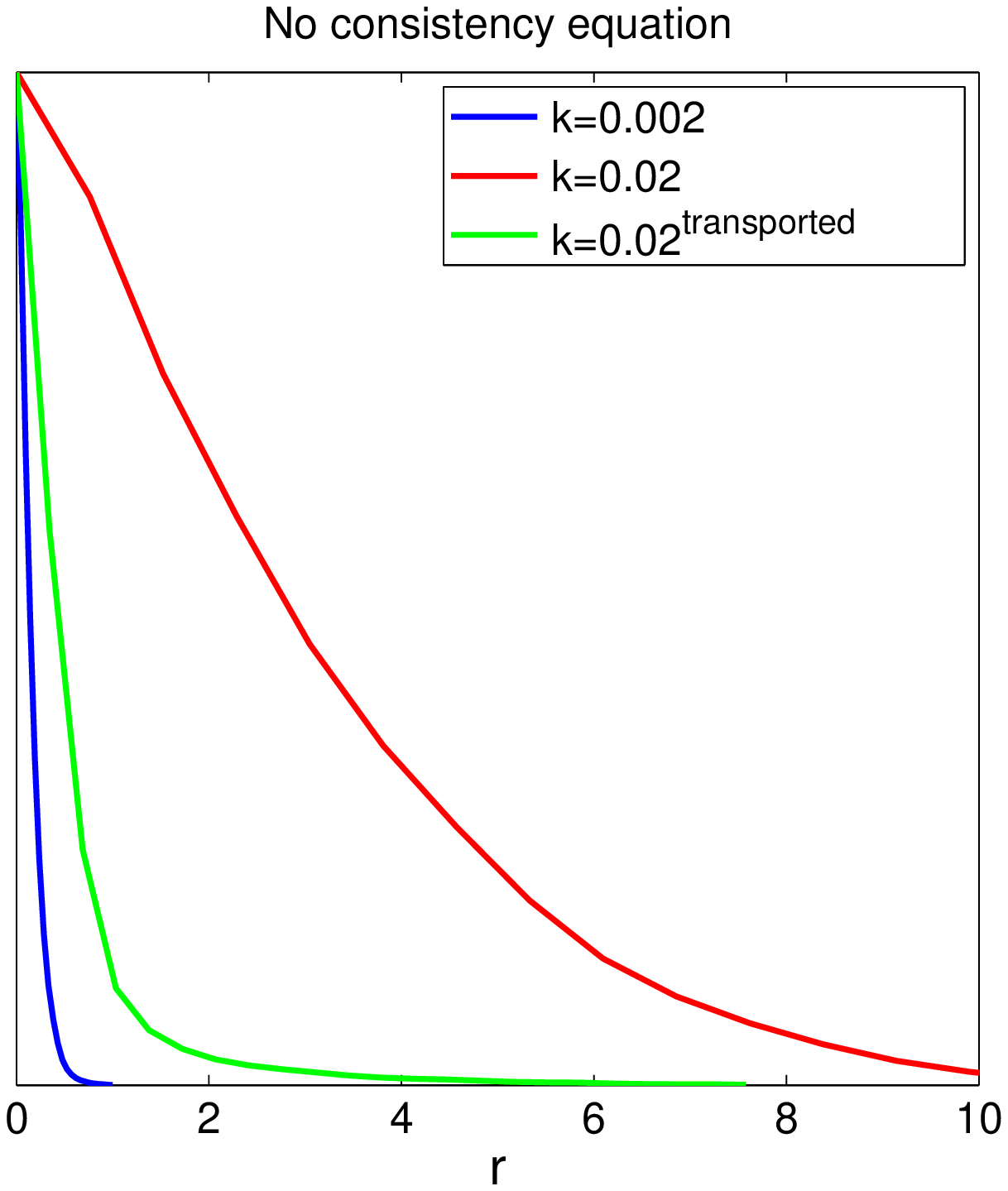}\\
\end{array}$
\caption{One-dimensional constraints on $r$. In the upper panels they
  are grouped by scale and in the lower by class of models.}
\label{1D_r} 
\end{figure*}

We now turn to examining constraints on $r$ under different
assumptions for the inflation model.  We assess the effect of imposing
the single-field equality, $r=-8 \nT$, and that of restricting to the
multi-field region, $r\lesssim -8 \nT$, obtained by clipping out from
the full case the models that don't obey the inequality. We then
obtain constraints for these models at different scales, and their
response to variations in the prior imposed at each scale.
Figure~\ref{1D_r} shows the one-dimensional marginalized constraints
on $r$ grouped in two different ways.

In the upper row of Fig.~\ref{1D_r} the grouping is by scale: the
first two panels show constraints obtained by sampling from the
likelihood under a uniform prior at $k=0.002 \impc$ and $k=0.02
\impc$, and the last column shows the chain transported from $k=0.002
\impc$ to $k=0.02 \impc$, which results in a non-uniform prior at the
new scale. We see that at the scale $k = 0.002 \impc$ used by WMAP,
the constraint is only modestly dependent on the prior. By constrast,
at $k=0.02 \impc$ the constraint on $r$ changes greatly under the
different prior assumptions, the centre panel showing the effect of
the choice of model type, and the right panel then showing further
modification when the alternative prior, induced from a uniform prior
at $k = 0.002 \impc$, is used instead.

The lower row shows the same results grouped by model type. In the
single-field case (left panel) the constraint on $r$ does not much
change when we alter the pivot scale or vary our prior. This is a
consequence of $\nT$ being less than zero by assumption, preventing
the tensors from growing towards shorter scales, in combination with
the observed near scale-invariance of the scalar spectrum across the
scales we are considering. As we have already seen, the data alone do
not significantly constrain the tensor amplitude at $0.02 \impc$,
without this additional model assumption.  Interestingly, sampling
under the consistency equation assumption at $k=0.002 \impc$ actually
gives weaker bounds on $r$ than the case of freely-varying $r$ and
$\nT$, as it happens that the models not satisfying the consistency
equation typically cannot fit the data with $r$ values as high as in
the single-field case.

The multi-field constraints on $r$ are given by the middle panel in
the bottom row and are tighter than in single-field. That is because
here we are sampling from the region {\em below} the consistency
equation line, which corresponds to smaller $r$, as in
Figs.~\ref{ntr0.002_ntr0.02}. Also, we see that the region
passing the multi-field inequality at the tensor pivot,
Fig.~\ref{ntr0.002_ntr0.02} (left panel), is much larger than the region 
at the smaller scale $k=0.02 \impc$ (right panel). At $k=0.002\impc$ 
about 50\% of the models are multi-field, against a scarce 2\% at 
$k=0.02\impc$. This also results in lower statistics at $k=0.02\impc$ in 
our procedure for selecting multi-field models from the full case, 
and accounts for the fluctuations seen in the middle bottom panel of 
Fig.~\ref{1D_r} (red line).  The same panel shows that constraints for 
multi-field models are, as in single-field, rather robust to changes in 
our prior. The variation here is more evident however: $\nT$ is more 
negative than in the single-field case, so going to smaller scales leads 
to smaller $r$ and tightens the contours as compared to the case at 
the tensor pivot (green, red and blue lines in bottom middle panel).

Overall, Fig.~\ref{1D_r} illustrates how constraints on tensor
quantities with present-day data are rather sensitive to one's choice
of prior and cosmological scale probed.  This means that at current
sensitivities, in addition to uncertainty in the data, one must admit
an uncertainty arising from our prior assumptions, for which the
variation we see in Fig.~\ref{1D_r} is an indication. This is not the
end of story, as priors leading to more extreme variations could be
envisaged too, and we have investigated only the transformation of a
uniform prior.

\section{Transformation of observables with cosmological scale}

We now return to the issue of choice of scale to impose constraints on
the tensors. This issue in fact splits into two separate ones:
\begin{enumerate}
\item On what scale is the prior probability distribution imposed (for
  example the choice of a uniform prior on some specific scale)?
\item For a given choice of prior, on what scale are the tensors
  optimally constrained?
\end{enumerate}

\subsection{Scale transformations and the choice of priors}

Our discussion above focussed on the first of these. Only
when parameters transform linearly with scale will the prior density
be preserved when transforming between cosmological scales $k$,
because linear transformations then amount to a rotation in the 2D plane 
of
the parameters considered.  This is the case for $\ln \AT$ and $\nS$,
for example, which transform linearly with scale,
\beqa
\ln \AT(k)&=&\ln \AT(k_0) +\nT(k_0) \ln \frac{k}{k_0}\,,\\
\nS(k)&=&\nS(k_0) + \frac{d \nS}{d\ln k}(k_0) \ln \frac{k}{k_0}\,.
\eeqa

For parameters that do not transform linearly this is not usually
true. In this case the density of models in parameter space is
modified as parameters do not keep their relative proportions when we
shift with $k$. This is what happens with $r$, where the transformation 
is exponential, depending on $\nS$, $\nT$ and scalar running
according to
\be
r(k)= r(k_0) \left(\frac{k}{k_0}\right) ^ {\nT(k_0) - \left
  [\nS(k_0)-1\right] - \alpha(k_0) 
  \ln{k/k_0}}.
 \label{r_transf}
\ee
As we saw in Fig.~\ref{ntr_prior_ovrl} (left panel), this can substantially 
modify 
the
prior relative to assuming a uniform prior at the new scale.  The net
effect on the prior in $r$, when going towards smaller scales, is a
compression in $r$ regions with negative $\nT$, raising its prior
density, and expansion of regions that correspond to positive $\nT$
lowering their prior density. If the parameters were well constrained
data could overcome this change in prior, but unfortunately they are
not. 

For our choice of smaller scale, $k=0.02 \impc$, this differentiated
sampling amounts to the posterior preferring a best-fit $\nT$
different by one-sigma from the one at $k=0.002 \impc$, as shown 
in Fig.~\ref{ntr0.002_ntr0.02}. We would expect an 
even larger variation in the posterior if we were to sample $r$ at 
smaller scales. This change is a consequence of the variation in prior 
alone, since we are not including any extra information when we 
sample at smaller scales.

From a purely theoretical point of view a prior imposed at one scale
is as plausible as that imposed at the next. Unless the theoretical
framework selects for a particular scale as preferred to specify
parameters at, our results for the significantly different constraints
on $r$ and $\nT$ at each scale are equally reasonable: a large
positive tilt of the tensor power spectrum is just as plausible as a
nearly scale invariant spectrum.

One might wonder whether sampling from a log prior on $r$, or $\AT$, 
would
help.\footnote{One might even suspect that $\AT$ is already
  effectively being sampled from a log prior by inheritting the prior
  on $\log(\AS)$ in combination with the uniform prior on $r$. But
  since $\AS$ is tightly constrained by data its prior is fairly flat
  in the region where the likelihood peaks, so the fact that $r$ has a
  uniform prior means that $\AT$ is effectively being sampled from a
  uniform prior as well.} However with a log prior any well-motivated
lower cut-off is likely to be at an extremely low value of $r$, and
already puts most of the prior parameter range well below future
observational sensitivity. While this doesn't prevent a detection, in
absence of a detection any observational limits are going to be
dominated by the prior rather than the data.

\subsection{Variation of constraints with pivot --- implications for 
inflationary models}

\begin{figure} [t]
\centering
\includegraphics[width= 0.7\textwidth]
{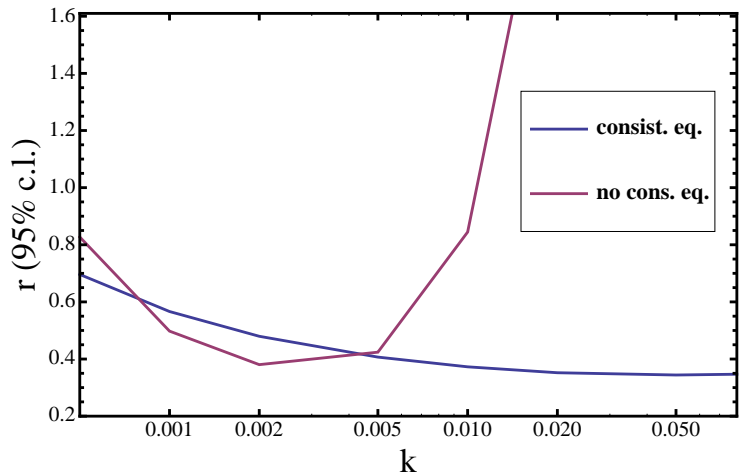}
\caption{Variation of the 95\% confidence limits on $r$ with scale. We
  compare the variation of chains which have the consistency equation
  imposed (blue), with those that have not (red). The prior on $r$ is
  uniform at $0.002 \impc$ in both cases.}
\label{r_areas} 
\end{figure}

We now turn to the question of the scale on which constraints should
be imposed once the prior is fixed. Figure~\ref{r_areas} shows the
change in 95\% confidence limits on the tensor-to-scalar ratio when we
probe at different $k$ scales. We compare this effect on chains which
have the consistency equation imposed with those that haven't.  

The non-inflationary runs have a quite well-defined pivot point,
outside of which constraints rapidly deteriorate. The strongest
constraints are obtained at $k=0.002 \impc$, with the WMAP constraints
on tensors coming mainly from the amplitude of TT modes at low
multipoles \cite{wmap5,wmap7}.  Due to cosmic variance the constraints
are dominated by statistical uncertainty up to multipoles $\ell\sim10$
corresponding to $k \sim0.002\impc$.  The upper limits quickly go up
as we move away from this pivot, towards smaller scales, and allow for
$r$ to become as large as 6 at $k\simeq 0.02\impc$.

However, for the consistency equation runs, the situation is very
different. Constraints on $r$ appear robust to variation with scale in
this case. This is because enforcing the consistency relation imposes
a strong correlation on $\nS$ and $r$, since it makes the assumption
these can be described jointly in terms of the slow-roll parameters,
given in Eq.~(\ref{epseta}) via Eq.~(\ref{nSnT}). This imposition
mimics apparent robust constraints on tensor quantities at all angular
scales, when in fact it is the prior rather than the data that is
ruling out sizable tensor mode contributions at all but the very
largest scales.

\begin{figure} [t]
\begin{center}
\includegraphics[width= 0.7\textwidth]{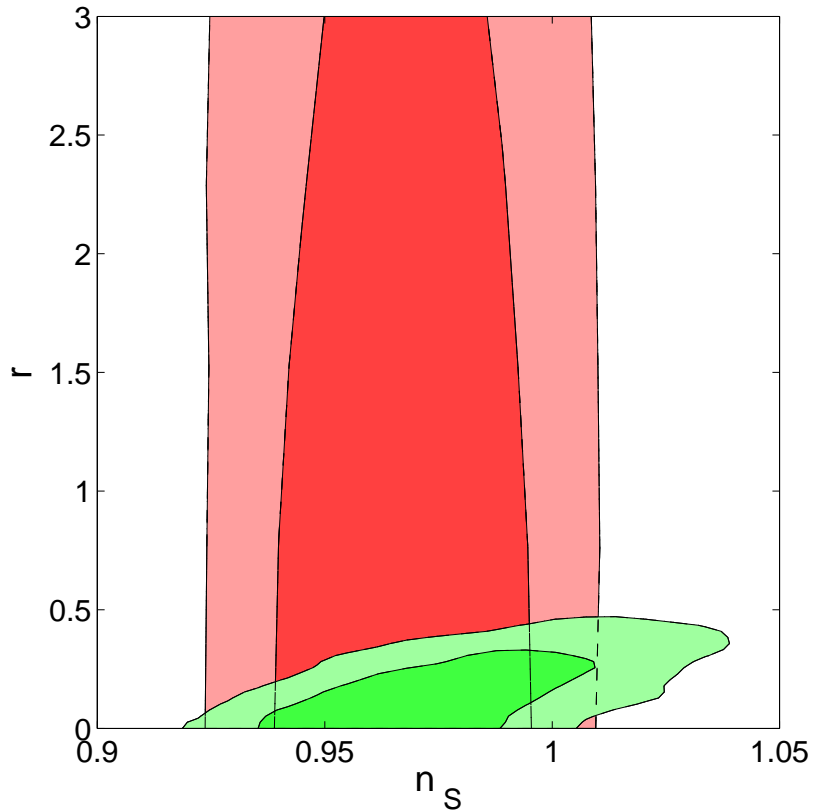}
\caption{An overlay of two $\nS$--$r$ distributions shown at
  $k=0.02\impc$, the scalar pivot scale. Green contours show
  constraints when the consistency equation is enforced and red when
  it is not. Relaxing the consistency equation prior decorrelates
  $\nS$ and $r$ and as a result, in the absence of the inflationary
  prior, $\nS$ actually becomes better determined.}
\label{nSr_overlay} 
\end{center}
\end{figure}

Enforcing the consistency equation on small scales, around
$k=0.02\impc$, excludes significant parameter space that would
otherwise be available in the $\nS$--$r$ plane, as shown in
Fig.~\ref{nSr_overlay}. It may be that some of this space is
available in models with non-inflationary sources for primordial
gravitational waves, or in the more general set of inflationary
proposals (multiple-field, higher-order corrections) that deviate from
single-field slow roll. These are expected to have their own internal
consistency relations that do not necessarily comply with the
single-field one.

\section{Conclusions}

The simplest models of single-field slow-roll inflation predict a
simple connection between the scalar and tensor contributions to the
spectra of perturbations, given by the consistency equation. Aside
from those there is a wide variety of models that predict deviations
from the simplest models, as well as non-inflationary proposals for
the origin of perturbations.

The tensor-to-scalar ratio, and therefore connections between the
scalar and tensor perturbations, are a powerful discriminator between
models.  We have shown that, with the current state of knowledge,
extraction of constraints on $r$ depends significantly on our prior
assumptions for the form of scalar and tensor perturbations.  We show
that enforcing the consistency relation leads to a reduction of
available parameter space by a factor of 10 or larger when quoting
constraints at the usual scales around $k=0.01\impc$ to $0.05\impc$
for the $\nS$--$r$ plane.  As a result, constraints obtained under the
assumption of this relation should be used for studying models of
single-field slow roll alone, for which it is valid. 
For other inflation proposals an analysis based on the imposition of 
the consistency equation can lead to artificial inferences.

In particular, the allowed values for the tensor-to-scalar ratio can
be significantly different from those one would expect from
traditional fits. Even when combining multiple datasets, data still
allows for tensor mode amplitudes which are several times larger than
the amplitude of scalar modes at scales around $k=0.02\impc$.
Furthermore, the ekyprotic- and collapse-type models prediction for
positive spectral index of tensor perturbations, $\nT>0$, is as valid
as the inflationary equivalent which predicts $\nT<0$.

We conclude that constraints on tensors presently have significant
prior dependence, and must be interpreted with care in light of the
particular models to be studied. Even the apparently innocuous
assumption of placing uniform priors at one scale rather than another
can significantly modify the constraints obtained, whether or not the
consistency equation or inequality is imposed.


\begin{acknowledgments}
We thank Eiichiro Komatsu and Antony Lewis for helpful discussions.
M.C.\ thanks the Astronomy Centre at the University of Sussex for
hospitality during this work.  M.C.\ was supported by the Director,
Office of Science, Office of High Energy Physics, of the
U.S. Department of Energy under Contract
No.\ DE-AC02-05CH11231. A.R.L.\ was supported by the Science and
Technology Facilities Council [grant numbers ST/F002858/1 and
  ST/I000976/1]. D.P.\ was supported by the Australian Research
Council through a Discovery Project grant.
\end{acknowledgments}


\end{document}